\RequirePackage{fix-cm}
\RequirePackage{hyperref}
\documentclass[smallextended]{svjour3}       % onecolumn (second format)
\smartqed  % flush right qed marks, e.g. at end of proof
\usepackage{graphicx}
\usepackage{verbatim}
\usepackage{natbib}
\usepackage{xcolor}

\usepackage{multirow}
\usepackage{booktabs}
\usepackage{longtable}
\usepackage{ragged2e}
\usepackage{todonotes}
\presetkeys%
    {todonotes}%
    {inline,backgroundcolor=yellow}{}

\journalname{Artificial Intelligence Review}
\usepackage[margin=3cm]{geometry}

\begin{document}

\title{Evaluating Conversational Recommender Systems: A Landscape of Research}
%\subtitle{}
\author{Dietmar Jannach }


\institute{D. Jannach\at
              University of Klagenfurt, Austria \\
              ORCID: 0000-0002-4698-8507 \\
              \email{dietmar.jannach@aau.at}           %  \\
}
\date{}

\maketitle

\begin{abstract}

Conversational recommender systems aim to interactively support online users in their information search and decision-making processes in an intuitive way. With the latest advances in voice-controlled devices, natural language processing, and AI in general, such systems received increased attention in recent years. Technically, conversational recommenders are usually complex multi-component applications and often consist of multiple machine learning models and a natural language user interface. Evaluating such a complex system in a holistic way can therefore be challenging, as it  requires \emph{(i)} the assessment of the quality of the different learning components, and \emph{(ii)} the quality perception of the system as a whole by users. Thus, a mixed methods approach is often required, which may combine objective (computational) and subjective (perception-oriented) evaluation techniques. In this paper, we review common evaluation approaches for conversational recommender systems, identify possible limitations, and outline future directions towards more holistic evaluation practices.

\end{abstract}

% =======================================
\section{Introduction}
\label{sec:introduction}
% =======================================
Personalized recommendations are a central part of many of today's popular websites and online services, where they can represent an important means to create value both for consumers and businesses~\citep{Gomez-Uribe:2015:NRS:2869770.2843948,jannachjugovactmis2019}. From the user interface (UI) perspective, recommendations are usually presented through dynamically filled and often personalized lists of items that are shown to users when they navigate the site or application.
In several online applications---including Netflix, YouTube or Amazon---such lists with item recommendations are even the most predominant interaction mechanism and occupy much of the space of the user interface.
From the interaction perspective, however, such approaches mostly implement a one-dimensional flow of information in the direction from the system to the user. While the system might record a user's reaction to a recommendation, e.g., when the user makes an action on a recommended item, users are typically not provided with means to express their particular thoughts about a specific item or to request the recommendation of an alternative item.

In various application domains, richer forms of information exchange between the system and the user may however be desirable. Think, for example, of someone seeking a recommendation from a friend for a restaurant to eat out this evening. In a real-world conversation, there might be several interaction turns between an information seeker and a person giving advice about restaurants, see also~\citep{christakopoulou2016towards}. The information seeker might, for example, first specify some initial preferences but may then also gradually revise them once she or he learns about the actually available space of options from the recommendation provider. Moreover, a recommendation seeker might want to hear an explanation why the friend recommends a certain restaurant.

Such application scenarios are usually not in the focus of today's ``one-shot'' recommender systems on e-commerce or media streaming sites. In this context, the promise of Conversational Recommender Systems (CRS)~\citep{jannach2021crscsur} is to fill this gap and provide such more natural and more interactive forms of online advice-giving. Interactive and conversational recommender systems have been discussed in the literature since the mid-1990s~\citep{hammond1994findme,DBLP:conf/aaai/BurkeHY96}. In recent years, we have however observed a newly increased interest in this area, which is fueled by a number of parallel developments. First, major advances were made in natural language processing (NLP) technology, for example, in the area of speech recognition and natural language understanding, which is crucial when the goal is to support naturally-feeling interactions, both voice-based ones or ones based on typed chatbot-like interactions. Second, along with the popularity of neural networks and machine learning in general, advances were made in other learning tasks that are often part of modern CRS, including the core task of determining suitable items for recommendation. Finally, we have also observed relevant developments in the hardware sector, and we are nowadays increasingly used to interact with electronic devices---in particular with smartphones and home assistants---in natural language.

Generally, \emph{building a CRS} can come with additional challenges that are not present in traditional, non-conversational recommenders. In many applications, there are challenges regarding natural language processing, e.g., to correctly understand user utterances. Moreover, a CRS must also perform some form of dialogue management and, for example, dynamically decide on the next conversational move after observing an action by the user. This multi-component nature of the system however also makes it more difficult to \emph{evaluate a CRS}. For example, when a user abandons an ongoing recommendation dialogue, there might be several reasons for it: the CRS might have had difficulties to process the user utterances, did not correctly recognize the user's specific intent, was not able to properly respond to an intent, or made recommendations that were not considered helpful.

In the academic literature, the evaluation of recommendation systems is largely focused on the underlying algorithms and specifically on their capability of predicting the relevance of items to individual users. Evaluating an entire \emph{system}---as opposed to an algorithm---however requires a much more comprehensive approach, as such an evaluation might not only involve recommendation quality factors beyond accuracy, e.g., diversity or novelty, but may also require us to assess user experience aspects or, ultimately, the impact of a system on its users~\citep{Herlocker:2004:ECF:963770.963772,Shani2015,jannach2021mcnamara}. Given that any CRS is a highly interactive system, it is in most cases not meaningful to evaluate the quality of the recommendations in isolation. Instead, a more holistic approach is needed, which in particular puts aspects of the \emph{users' perceptions} of the system in the center of the evaluation.

In the context of traditional recommender systems, various \emph{quasi}-standards emerged over the years, in particular with respect to \emph{offline} evaluation procedures using historical datasets, see~\citep{Herlocker:2004:ECF:963770.963772,Shani2015}. Furthermore, a few frameworks for \emph{user-centric} evaluation of recommenders are around for about a decade~\citep{Pu:2011:UEF:2043932.2043962,umuai2012knijnenburg}. No common standards however exist yet for the area of conversational recommender systems and it is not even entirely clear which evaluation approaches are commonly taken in the literature. With this paper, we aim to narrow this research gap by providing a survey of existing approaches to the evaluation of CRS. The manuscript has both elements of a survey as it reports what has been applied in the literature; and it has elements of a tutorial, as it provides a starting point for researchers on the opportunities and challenges regarding the evaluation of conversational recommenders.

The paper is organized as follows. Next, in Section~\ref{sec:conceptual-architecture}, we review a typical conceptual architecture of a CRS. Afterwards, in Section~\ref{sec:evaluation-methods}, we discuss the different paradigms of evaluating a CRS. In Section~\ref{sec:landscape}, we then paint a landscape of existing research on CRS in terms of domains, evaluation paradigms and specific evaluation measures. A discussion of our observations and research implications are finally provided in Section~\ref{sec:discussion}.

\section{Conceptual Architecture of a CRS}
\label{sec:conceptual-architecture}

In this section, we first review typical components of CRS on a conceptual level (Section~\ref{subsec:components}) and then discuss the relationship of CRS to general task-oriented dialogue systems (Section~\ref{subsec:relation-to-task-oriented-systems}).

\subsection{Components of a CRS}
\label{subsec:components}
\begin{figure}[h!t]
  \centering
  \includegraphics[width=0.6\linewidth]{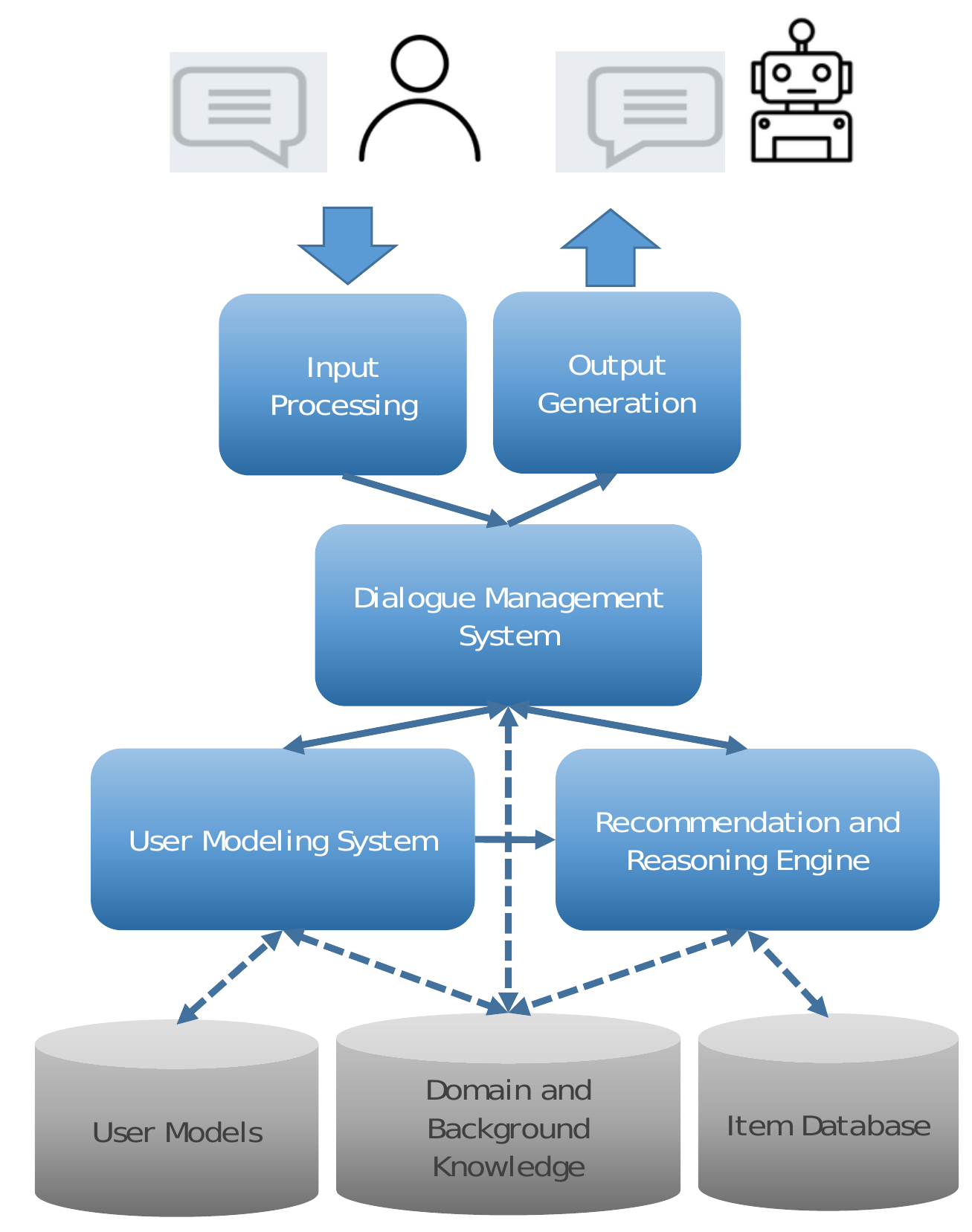}
  \caption{Common architecture of a conversational recommender system (see also~\cite{Thompson:2004:PSC:1622467.1622479,jannach2021crscsur})}.
  \label{fig:architecture}
  %\vspace{-6pt}
\end{figure}

Figure~\ref{fig:architecture} shows the typical conceptual components of the architecture of a CRS, illustrated for a system that supports natural language interactions. Note that various CRS were proposed in the literature that are based on structured web interfaces using forms and buttons, in particular ones based on critiquing~\citep{chenpucritiquing2012}. In this work, we mainly focus on current and future systems that support natural language inputs and outputs. Many challenges regarding the evaluation of such systems are however independent of the interaction modality.

In the sketched architecture, at each interaction cycle the first task of the system is to process the actions (here: utterances) by the user. This processing phase often consists of voice-to-text conversion in a first step. Afterwards, a Natural Language Understanding (NLU) processing chain may be invoked, which may, for example, include a Named Entity Recognition (NER) step, the recognition of the user \emph{intent} (e.g., providing a preference or asking for an explanation), and the extraction of attributes from the utterance that relate to this intent. Overall, the Input Processing module takes the user's actions (e.g., menu commands, typed or spoken language, or gestures) as an input and analyzes it in different dimensions to produce a structured output that can be further processed (e.g., extracted keywords, preference statements, identified entities, or assumed user intent). Let us iterate here that the architecture in Figure~\ref{fig:architecture} is a conceptual one. In a particular technical implementation, several software modules might exist to implement a conceptual module. For example, voice-to-text translation and intent recognition might most often be separate components of an Input Processing module.

The outputs of the Input Processing phase are then processed and used by the Dialogue Management System (DMS), which can be seen as the central (conceptual) component of any CRS. One of its main tasks is to decide on the next action by the system, often based on domain-specific information and background knowledge. In case new preference information has become available in the current interaction, the DMS will forward this information to the User Modeling System (UMS). The UMS takes this information to update the corresponding user model in its database. For modeling a user, the UMS may also consider domain information or background information (e.g., about relevant user characteristics in a domain).

If the DMS then decides to present a recommendation as a next conversational move, it will invoke the underlying recommendation component to retrieve a set of item suggestions to present to the current user. The Recommendation and Reasoning Engine in this case receives a user model from the UMS and then determines a set of suitable items based on the given catalog and maybe some background knowledge (e.g., about contextual factors like time or based on domain-specific recommendation rules). If, on the other hand, other conversational moves seem more appropriate, the DMS might use other reasoning components to, e.g., provide an explanation for the previous recommendation or to decide which question to ask next to the user. Again, this reasoning is often based on domain-specific knowledge.

Once an appropriate response is determined by the DMS with the help of the Recommendation and Reasoning Engine, the information is forwarded to the Output Processing Module. This module then takes the necessary steps to generate outputs that are presented to the user. Depending on the given modality, the final output might for example be generated by invoking a text-to-speech processing module that reads out a recommendation to the user.

Internally, a CRS can rely on a variety of knowledge sources or databases, as mentioned above. Clearly, like any recommender system, the CRS has to have a database of recommendable items. Additional sources of knowledge may include \emph{``world knowledge''} for the named entity recognition task, meta-data about the features of the recommendable items, a pre-defined list of supported user intents~\citep{DBLP:conf/um/CaiC20}, \emph{dialogue knowledge} in the form of possible conversation states and transitions between the states and so forth.

For several of these components machine learning models are commonly used. Like in traditional non-conversational recommender systems, one central machine learning model is used to make personalized item suggestions based on existing user preferences and other types of information like community ratings. In CRS, additional models can be trained, for example for the individual input processing tasks or for deciding on the next conversational move. In this context, a particular integrated approach is taken by ``end-to-end'' learning conversational recommender systems. In such systems, the idea is to learn how to respond to a user utterance by training a machine learning model using a larger collection of previously recorded recommendation dialogues. Typical examples of such systems are the deep learning-based system presented by~\cite{li2018towards} or the KBRD and KGSF systems by~\cite{chen-etal-2019-towards} and~\cite{kgsf2020}. Both approaches are based on the ReDial dataset~\citep{li2018towards}, which consists of several thousand movie recommendation dialogues between human information seekers and a human recommender. A particularity of the dataset is that it was created with the help of crowdworkers, which, as we will see, can lead to a number of challenges.

Overall, we observe that the output that is delivered by a CRS and the perceived quality of this output may depend on various factors. In fact, what makes the task of building and evaluating a CRS difficult is that there are many places where the system can fail. In a traditional recommender system, the failure might mainly be that the item suggestions are not very relevant. In a CRS, the system might in addition already fail to properly understand the utterance or misinterpret the user's current intent. Likewise, in an end-to-end learning system that uses natural language generation, the CRS might produce repetitive or even ungrammatical sentences at the output side. From the perspective of the evaluation of a CRS, there are therefore various components and machine learning models that one might have to examine with respect to their performance, as they might all contribute to the resulting user experience. In addition, it is important to evaluate the system's quality level as a whole by looking at the responses it returns during an entire conversation.

\subsection{Relation to Task-Oriented Dialogue Systems}
\label{subsec:relation-to-task-oriented-systems}
We note here that the described CRS architecture has various elements of the more general class of \emph{task-oriented} dialogue systems. Such systems are nowadays typically able to support natural language interactions (e.g., with voice input and output) and may support both simple, one-shot tasks---like playing music of a given artist---as well as more complex tasks like a flight reservation. The architecture of such systems therefore usually includes components for Natural Language Understanding, intent classification, as well as Named Entity Recognition~\citep{Chen2017SurveyonDialog,Gao2018NeuralApproaches}. Likewise, for more complex tasks, they often have components for state tracking or ``slot filling''~\citep{balaraman-etal-2021-recent,LouvanM20Recent}. Moreover, end-to-end learning approaches are nowadays very common for task-oriented systems, as is the case for CRS.

Given these commonalities, CRS can be seen as a subclass of task-oriented dialogue systems, which are focused on the specific tasks of preference elicitation, item recommendation, and sometimes explanation. CRS are however not necessarily based on natural language interaction, consider, e.g., traditional critiquing and form-based advisory systems~\citep{chenpucritiquing2012,Jannach:2004:ASK:3000001.3000153}. In terms of evaluation, however, various approaches from task-oriented dialogue systems can be used for the evaluation of CRS as well, see~\citep{finch-choi-2020-towards} for an overview. Still, the evaluation of CRS also requires task-specific approaches, e.g., to gauge the quality of the recommendations.

\section{Evaluation Paradigms for Conversational Recommenders}
The goal of this section is to introduce the \emph{general} methodological approaches to evaluate a CRS.
Furthermore, we review common quality dimensions when evaluating a CRS in a broader view based on the categorization by \cite{jannach2021crscsur}.
Later on, in Section~\ref{sec:landscape}, we provide a more detailed landscape of today's common research practices based on the reviewed literature. This landscape may then also help us identify potential gaps of current research.

\subsection{Evaluation Paradigms}
\label{sec:evaluation-methods}
In research on traditional recommender systems, we usually find three main types of general paradigms, see also~\citep{Herlocker:2004:ECF:963770.963772,Shani2015}:

\begin{enumerate}
  \item Field studies, commonly in the form of A/B tests, where a system is evaluated in a real-world environment.
  \item User-centric research, often in the form of controlled experiments, involving humans interacting with a system (prototype) developed for a user study.
  \item Computational studies that do not involve humans, commonly used to assess the prediction accuracy of algorithms or other properties of computational elements of a CRS, commonly through ``offline experiments''.
\end{enumerate}

\paragraph{Field Studies.}
In field studies, the effects of a real-world deployed system are analyzed, typically focusing on business-oriented key performance indicators for the business value of the system such as sales, revenue, or engagement~\citep{jannachjugovactmis2019}. Field studies can take different forms. The most informative form of a field study is a controlled experiment (often called A/B test). Here, two or more versions of a deployed system are created, e.g., two versions that have a slightly different user interface. Then, users are randomly assigned to one of these systems versions, forming \emph{treatment} and \emph{control} groups. After some defined period, e.g., a few weeks, potential differences in the behavior of the two user groups are analyzed, e.g., in terms of their engagement with the system or in terms of the generated revenue. Since in such controlled experiments only one single \emph{independent} variable is changed between the groups and everything else is kept constant, any observed difference in the outcome, i.e., in the \emph{dependent} variable(s), are attributed to this change.

Other forms of field studies exist as well. \emph{Quasi-experiments} are very similar to controlled experiments, except that the assignment to treatment and control groups is not randomized. An example of such type of research can be found in~\citep{Delgado2002KnowledgeBA}, where the behavior of visitors of a tourism website was compared by dividing them into a group of people who interacted with a recommender and those who did not. Besides such experimental studies, also \emph{observational} studies can be made, where for example longitudinal effects of a recommender system are observed. In~\citep{ZankerBricmanEtAl2006}, for example, the sales distribution in an online shop was monitored over time after the implementation of a CRS. Finally, surveys among real users are also a common instrument in practice to assess a fielded system. Differently, e.g., from A/B tests, such surveys focus on subjective quality experiences of users and not on aspects that can be objectively determined such as conversion rates or clicks.

Field studies and in particular A/B tests are clearly the most informative type of evaluation approach, as they assess a system in its context of use. Running such A/B tests however also comes with challenges, including the effort to build various system versions and the risk of deploying a system that hurts the user experience, potentially leading to loss of consumer trust. See ~\citep{KohaviAB2020} for an in-depth discussion of how to design, run and evaluate A/B tests in practice. Furthermore, see~\citep{Hofman2016}, who review techniques for \emph{online evaluation} of information retrieval systems, which in general share a number of commonalities with recommender systems.

\paragraph{User-Centric Research.} Given the costs and risks of field tests and given that academic researchers usually do not have access to deployed systems, studies involving humans, e.g, in a laboratory environment, represent an important alternative to evaluate certain aspects of a recommender system. Again, controlled experiments are one of the most informative forms of such studies. Like in A/B tests, users are confronted with two or more versions of a system, where usually only one independent variable is changed to observe its effects on one or more dependent variables. Differently from A/B tests, the compared systems are not actually deployed ones and often created for the purpose of the study. Correspondingly, the study participants (or: subjects) are not real users, but often students, subjects that were involved through \emph{convenience sampling}\footnote{Convenience sampling involves subjects that are easy to reach and can be appropriate for pilot testing.}, or subjects that are recruited with the help of professional services or crowdsourcing platforms like Amazon Mechanical Turk.

In studies involving users, both \emph{objective} and \emph{subjective} measurements can be made. Objective measures could for example include the \emph{time} a subject needs to make a decision (supported by a CRS). A corresponding subjective measure could be made by asking the participants after interacting with the system through a questionnaire about their \emph{perceived effort} to make a decision.

Besides controlled experiments, also other forms of user-centric research exist. In particular in CRS research, individual \emph{human judges} are commonly involved to assess the quality of the individual responses (utterances) generated by a system in a given dialogue situation. Such assessments can be done both on an absolute and a relative scale and various quality dimensions can be considered, e.g., the consistency of the generated utterance with the ongoing conversation.
Moreover, various types of \emph{qualitative research} methods, including for example interviews or focus groups, are widely used in practice and in other academic fields outside of recommender systems to gather a deeper understanding of a given problem. Such forms of research can for example be used to understand the needs and behavioral patterns of recommendation seekers in human-to-human conversations. % cite some radlinski work here?
Finally, researchers in CRS and dialogue systems in general often use \emph{case studies} to demonstrate the usefulness of their approach. These case studies often come in the form of example dialogues between the CRS and the user, and they are usually evaluated and hand-selected by the researcher.

Generally, user-centric research can provide us with many insights without the need of having access to a real-world system. Ultimately, since CRS are interactive systems, aspects related to the user experience of a CRS can only be reliably investigated by involving human subjects. Certain limitations of such studies however remain and must be kept in mind, e.g., that study participants might not be entirely representative of a real user population or that they might have different motivations in the typically artificial setting of a user study. An in-depth review of methods for evaluating \emph{interactive information retrieval} systems---which are similar to CRS in various ways---can be found in~\citep{Kelly2009Methods}.

\paragraph{Computational Studies.} Finally, various aspects of a CRS or of its individual components can be evaluated with the help of \emph{computational experiments}, which do not involve human subjects. Such types of experiments and analyses represent the most common research instrument in general recommender systems research.

Such studies predominantly aim to assess the ability of different recommendation \emph{algorithms} (or: machine learning models) to learn the relevance of individual items for individual users. Common evaluation procedures from machine learning are applied in that context to compare the relevance prediction or ranking accuracy of the algorithms. In the traditional \emph{rating prediction} task, for example, algorithms are compared by measuring their ability to correctly predict held-out test data, using metrics such as the Root Mean Squared Error (RMSE). Besides prediction accuracy, also other potential quality factors of recommendations can be assessed \emph{offline}, for example, the diversity of the generated recommendation lists, the general popularity of the recommendations, the novelty of the suggested items compared to those items a user has liked in the past, or the extent to which an algorithm has a tendency to concentrate the recommendations on a certain part of the catalog~\citep{Jannach2015umuaiWhat}.

Since a CRS commonly has an underlying algorithm to determine suitable recommendations for an ongoing conversation, such forms of evaluations can also be applied in the context of CRS. Differently from traditional RS research, such accuracy evaluations typically cover only one aspect of a more comprehensive evaluation of a CRS. However, similar hide-and-predict evaluation approaches are commonly used also for other components of a CRS, e.g., for determining the accuracy of an entity recognition module. A fundamental requirement in such offline evaluations is the availability of suitable datasets with ground-truth information, e.g., about the preferences of the users. The field of Information Retrieval (IR) has a long history of evaluating IR systems with the help of \emph{test collections} of documents with annotated ``ground-truth'', see~\citep{Sanderson2010Test} for an overview. However, a common problem when applying such an approach in the field of recommender systems is that for most of the items in the catalog no ground-truth is available, as users commonly only interact with a tiny fraction of the items.

Other types of offline studies can be found in the literature as well. Some works like proposed by~\cite{Gedas2019Longitudinal} or~\cite{ferrarojannachserra2020recsys} aim to understand \emph{longitudinal} effects of recommender systems, using (agent-based) \emph{simulation} as a research instrument, where the agents typically include the recommender system and the community of users. Such studies, e.g., on emerging effects of different recommendation strategies, are however uncommon in the context of CRS. Simulating the behavior of \emph{individual} (hypothetical) users who interact with the designed CRS is, on the other hand, rather common. In such cases, a rationally acting user is assumed who, for example, truthfully answers questions by the CRS about her preferences or if a presented recommendation matches her needs. In such simulations, one can for example assess the number of preference eliciting interactions that are needed until the CRS makes a suitable recommendation.

Finally, a number of computational analyses can be made with respect to \emph{linguistic properties} of the utterances made by a CRS that supports the generation of natural-language responses. In this context, we can for example compare the \emph{fluency} or \emph{perplexity} of the responses by different CRS implementations.

\paragraph{Discussion}
Ultimately, field studies are the most informative way of evaluating a CRS. In such studies, the evaluation is done in the system's context of use and measures regarding the utility of the CRS both for consumers and providers can be taken. Given the complexity and risks that comes with such field studies, researchers both in academia and industry therefore rely on alternative approaches deemed suitable to help to shed light on particular aspects of a CRS. User studies in particular may help to gauge subjective quality perceptions of a CRS in a controlled environment; computational experiments, on the other hand, are often considered useful to rule out some alternative recommendation algorithms due to their limited prediction performance.

Given the interactive nature of CRS, as compared to traditional recommender systems, a multi-method approach seems generally required when trying to assess the quality of a CRS in a comprehensive and informative way. While evaluations that exclusively rely on offline experimentation are very common in the literature, they carry the danger that the computational metrics that are used as proxies, e.g., the RMSE, are not indicative of or correlated with the usefulness and success of the system in practice~\citep{jannach2021mcnamara}. Likewise, in the context of CRS, it is difficult to know if higher values for a particular linguistic scoring method, e.g., the BLEU score, would always correspond with human quality perceptions~\citep{liu-etal-2016-evaluate}.

\subsection{Quality Dimensions of Conversational Recommenders}
\label{subsec:quality-dimensions}
There are various quality dimensions that are common to almost all sorts of recommender systems, including non-conversational and conversational ones. In-depth surveys on the evaluation of recommender systems in general can be found in~\citep{Herlocker:2004:ECF:963770.963772} and~\citep{Shani2015}. Typical quality dimensions for recommendations in the academic literature for example include ~\citep{Shani2015}:
\begin{itemize}
  \item Prediction accuracy, i.e., how good a recommender system is in identifying items that are relevant for users;
  \item Item coverage, i.e., how many items of the catalog are recommended by the system;
  \item Novelty, i.e., the capability of a system to help users discover new content;
  \item Serendipity, i.e., the ability of a system to recommend surprising, yet relevant content;
  \item Diversity, i.e., the capability of a system to create diversified, yet relevant recommendation lists;
\end{itemize}
In practical environments, various other metrics are used, mainly to assess the \emph{value} or \emph{utility} that a deployed system creates for its stakeholders, including, for example, increased sales or customer retention, see~\citep{jannach2021mcnamara} for an overview.

Beyond such general quality dimensions, there are also quality dimensions that are very specific to CRS and are, for example, related to user interaction aspects. On a general level, we can differentiate between the following quality dimensions~\citep{jannach2021crscsur}:
\begin{itemize}
  \item Effectiveness of Task Support
  \item Efficiency of Task Support
  \item Quality of the Conversation and Usability
  \item Effectiveness of Subtask
\end{itemize}

In the following, we briefly summarize these dimensions and provide examples for commonly used subjective and objective measures. For the following discussion, remember that various combinations of evaluation paradigm (field study, user-centric research, computational experiments) and quality dimensions are possible. A detailed analysis of current research practices is given afterwards in Section~\ref{sec:landscape}.

\subsubsection{Effectiveness of Task Support}
In general, any recommender system is designed to serve one or more purposes and create value for different stakeholders~\citep{JannachAdomavicius2016purpose,abdollahpouri2020}.
The large majority of academic research focuses on the \emph{value for consumers} and in particular how a recommender system supports its users during information search and decision-making tasks. In many cases, the underlying assumption is that increasing the value for consumers will at least indirectly lead to more value for the recommendation provider.

Ultimately, the relevant question in this context is if a CRS is able to create any \emph{utility} for the users. In a real-world deployment, different indicators can be considered. For example, we may assume that a system is useful when consumers \emph{repeatedly use} the system; or when they often \emph{accept the recommendations}, i.e., they adopt the suggestions by the system.
In academic environments, researchers usually do not have access to a real-world system,
and they thus rely on alternative research methodologies as discussed in the previous section.

In user-centric studies, one can for example \emph{objectively} measure the fraction of subjects who completed the given advice-seeking task or the fraction of users who accepted one of the recommendations made by the CRS under investigation. At the same time, one can ask study participants about their \emph{subjective} perceptions, for example, if they found the CRS useful for decision-making, how confident they are in their decision, or if they would use a similar system in the future.

In most user-centric studies, the focus is therefore on the consumer value of a CRS and the implicit assumption is that a CRS that creates value for the consumer will be beneficial from a business perspective as well. Generally, it is however also possible to investigate user behavior in a way that is more directly targeting at business value, see for example the study by~\cite{DBLP:journals/isr/AdomaviciusBCZ18} on the subjects' \emph{willingness-to-pay} when interacting with a recommender system.

In many computational studies on recommender systems in general, prediction accuracy is used as a proxy to assess how effective a system is in supporting its users. In CRS research, the same approach is often taken to evaluate the core recommendation algorithm as one part of an often more comprehensive evaluation.

\subsubsection{Efficiency of Task Support}
In many research works, the efficiency of the recommendation processes is considered a key quality factor of a CRS. Typically, the assumption is that a system is better if it supports its users to make decisions faster or, more generally, with less effort. Efficiency can be evaluated in a number of ways, using either research approaches that involve humans or computational studies. Also, efficiency can both be determined with the help of objective and subjective measures.

In studies involving users, i.e., in field studies and lab experiments, one can for example measure the average time users need to find an item they like or the time until they give up before finding such an item. Likewise, one can count the average number of interaction cycles before a recommendation is accepted. In terms of subjective measures, one can explicitly ask users about their perceptions regarding the effort that was needed to find a suitable item.

Efficiency is however also frequently assessed through the specific type of computational studies mentioned above, where a rationally-behaving user with pre-defined preferences is emulated. In such simulation studies, the CRS interacts with this simulated user---typically by asking questions about preferences and by presenting recommendations---and then the number of interaction cycles is counted until an item is identified by the CRS that matches the preferences of the simulated user. Offline evaluations of this type are common in the literature, when the technical proposal is related to the interaction strategy of the CRS, i.e., when the system is assumed to dynamically determine the next conversational move, e.g., through reinforcement learning.

\subsubsection{Quality of the Conversation and Usability}
A number of evaluations in the literature on CRS specifically focus on certain aspects related to conversation quality and to system usability in general. Conversation quality typically plays a major role in systems that interact with users in natural language.

Usability aspects are most commonly evaluated with user-centric research designs, e.g., by asking study participants to fill out a questionnaire with usability-related items. These questionnaires can either be based on standardized instruments such as the System Usability Scale (SUS) or using questionnaire items that were specifically designed for the purpose of the study. The items of such usability questionnaires for example comprise questions related to the ease-of-use of the system, its consistency, the users' feeling of control, or the intention of the study participants to use this or a similar system again in the future.  Some usability aspects can also be assessed objectively and though computational experiments. A typical objective measure that is related to usability are the response times, i.e., the time needed by the system to generate the next dialogue utterance. In addition, other proxy measures that are considered relevant for the system's usability can be used. On can, for example, assume that the needed user effort influences a system's usability and then rely on the above efficiency measures as usability indicators.

Various factors of the quality of the conversation can be assessed through user studies and questionnaire as well. One can, for example, ask the study participants about their perception of the consistency of the system responses with respect to the previous dialogue acts. Or, one can ask questions about the linguistic quality or understandability of the utterances returned by the system. Certain objective measurements can also be made during a user-centric study, e.g., by counting how often the CRS did not recognize the intent of the user's utterance correctly.

Finally, a few aspects regarding the quality of the conversation can also be investigated through computational studies.
In particular linguistic properties are often assessed also through computational measures, e.g., for fluency or diversity, as mentioned above.

\subsubsection{Effectiveness of Subtask}
A CRS, as mentioned above, usually is a multi-component, complex system. A number of research works therefore focus on evaluating how good individual components of the system fulfill their specific tasks. A typical example is to assess the success or failure rate of the input processing system, e.g., the entity recognizer, or the accuracy of the intent prediction system.

Again, both user-centric studies as well as computational experiments are possible. For example, in a study where participants interact with a CRS, one can record the dialogues and later manually analyze the system's success of correctly recognizing the users' intent or how often a user accepted a proposal to state certain preferences. In offline studies, on the other hand, one could create a dataset containing recorded interactions between an information seeker and a recommendation provider, label the seeker's utterances with their intent, and then train and evaluate machine learning models for intent classification.

\section{A Landscape of  Research}
\label{sec:landscape}
In this section, we review the research activity and practices in the field of CRS over the years, with the goal of identifying trends and potential research gaps. In particular we will focus on evaluation aspects and on the developments in the last few years, where we observed a significant trend towards natural language interfaces and deep learning models. Our analyses are based on a corpus of papers that was collected for the recent survey by~\cite{jannach2021crscsur}. The corpus contains almost 150 research papers that were identified through a semi-systematic literature\footnote{Papers were identified by querying various digital libraries and applying a snowballing procedure afterwards. The considered libraries included Springer Link, the ACM Digital Library, IEEE Xplore, ScienceDirect, arXiv.org, and ResearchGate. Search terms for example included ``conversational recommender system'',
``interactive recommendation'', ``advisory system'', or ``chatbot recommender''. } search and which were manually categorized by researchers.

We identified 127 papers in this corpus which we considered relevant for our analyses. Specifically, since our focus is on evaluation aspects, we did not consider papers that did not propose a method or tool, which could be evaluated. Therefore, we do not include papers of different types in the subsequent analysis, for example: pure survey works, papers that only discuss general challenges, works that present and analyze datasets, or exploratory studies on how humans interact in recommendation scenarios.\footnote{Note that the literature search might be incomplete to a certain extent, potentially missing works that were not listed or identified in the digital libraries. Despite this potential incompleteness, we are confident that the identified papers and the reported findings are representative for the research field.}

\begin{figure}[h!t]
  \centering
  \includegraphics[width=0.8\linewidth]{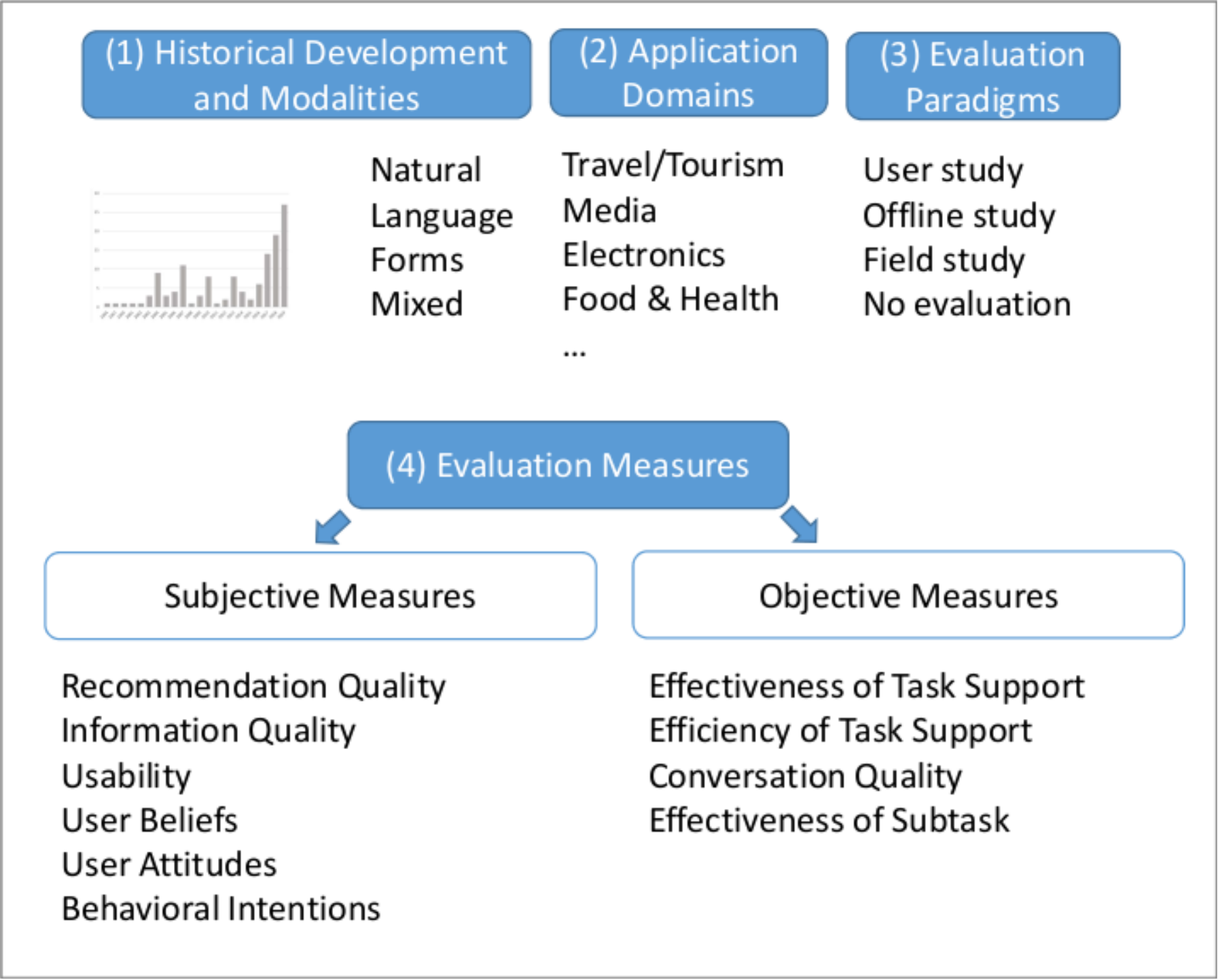}
  \caption{Landscape of CRS research discussed in Section~\ref{sec:landscape}.}
  \label{fig:landscape}
  %\vspace{-6pt}
\end{figure}

Figure~\ref{fig:landscape} provides an overview of the aspects that are covered in this section.

\subsection{Historical Developments: From Forms to Natural Language Interactions}
\begin{figure}[h!t]
  \centering
  \includegraphics[width=0.85\linewidth]{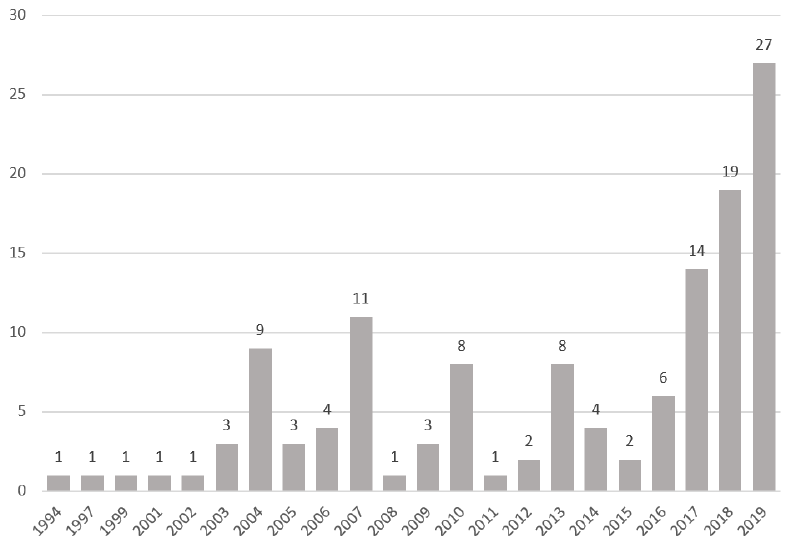}
  \caption{Number of papers published per year.}
  \label{fig:landscape-papers-per-year}
\end{figure}

Figure~\ref{fig:landscape-papers-per-year} shows the papers organized by publication year. In the 1990s, during the initial spread of the World Wide Web, only a few works were identified, which we would consider as predecessor of today's CRS, e.g., \citep{hammond1994findme,Burke:1999:WPS:315149.315486}. In the fifteen years between 2000 and 2015, interest in CRS was higher, but seemed to mostly stagnate. A strong increase in research interest is however observed starting around 2016, which was also the time where chatbots became popular and deep learning became the method of choice in algorithms research in recommender systems, e.g., \citep{qiu2017alime,Argal2018}.

\begin{figure}[h!t]
  \centering
  \includegraphics[width=0.65\linewidth]{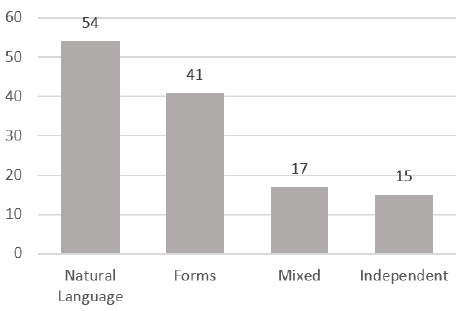}
  \caption{Modalities addressed in papers.}
  \label{fig:landscape-modalities}
\end{figure}

Figure~\ref{fig:landscape-modalities} illustrates which interaction modalities the surveyed papers target at. Overall, the largest number of papers are found on CRS that support natural language (NL) interaction, either in written and spoken form, e.g.,~\citep{chen-etal-2019-towards,zhou2020improving}. User interfaces based on forms and buttons, in particular ones that implement critiquing techniques or knowledge-based advisor approaches, see, e.g.,~\citep{Jannach:2004:ASK:3000001.3000153}, historically represent the second most frequent category.

A smaller number of the papers focuses on user interfaces that support mixed modalities. In the identified papers that use such mixed modalities, we often observe that one main modality (forms or NL) is complemented with an additional one, e.g., maps, a 3D space, or body gestures, e.g.,~\citep{MapBasedRicci2010,DECAROLIS201787}. Interestingly, only a handful of papers support NL input and output with form-based interactions, e.g.~\citep{IOVINE2020113250,Grasch:2013:RTC:2507157.2507161}. This is to some extent surprising as many chatbot platforms support both typed natural language input and output as well as buttons and other interaction elements, e.g., to select from pre-determined options.

Generally, the CRS we found were designed to run as apps on smartphones or desktop computers. Only in a few cases alternative modalities were in the focus, e.g., in the form of physical robots or interactive screens in brick-and-mortar stores.

A number of works was categorized as being \emph{independent} of the interaction modality. This collection of works for example includes papers on algorithmic improvements for deciding on the next conversational move, e.g., which critique to show to users, which could be implemented both in natural language or with forms, e.g.,~\citep{smyth2004compound}.

Finally, looking closer at the developments over time, we found that the huge majority of papers published in the last three years of our analysis, i.e., those between 2017-2019, aimed at natural language interfaces. Specifically, 49 of the 59 papers (83\%) published in these years were involving natural language interactions as their main modality. This trend is clearly fueled by the ongoing boom in machine learning in general and the availability of datasets containing human-human dialogues that can be used for training.

\subsection{Evaluation: Domains}
\begin{figure}[h!t]
  \centering
  \includegraphics[width=0.75\linewidth]{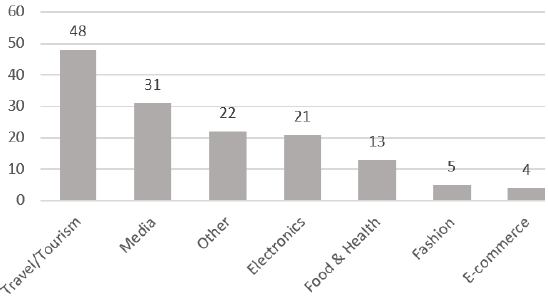}
  \caption{Domains addressed in papers.}
  \label{fig:landscape-domains}
\end{figure}

CRS can be applied in various domains. Figure~\ref{fig:landscape-domains} provides an overview of the domains addressed in the investigated papers. For this analysis, we have organized the target domains into a number of broader categories. In the majority of cases, only one application domain was in the focus of the research. Only in ten papers more than one domain was considered, e.g.,~\citep{smyth2004compound,Llorente2012} or more recently~\citep{Wu2019}. These papers are therefore counted more than once in Figure~\ref{fig:landscape-domains}.

Recommendation in the travel and tourism domain was most frequently researched in the examined papers. The typical items to recommend include destinations, points of interest (POIs) or restaurants, e.g.,~\citep{AverjanovaMobyRec2008,christakopoulou2016towards}, which we also subsumed in this category even though eating out is not always necessarily tied to traveling. The second most frequent category is the recommendation of media, including in particular movies, music, and news. Within this category, movie recommendation scenarios are predominant, which might at least partially be attributed to the existence of datasets containing recorded human-to-human conversations about movies. Among the other application domains, electronics (e.g., digital cameras), fashion, general e-commerce and health \& food recommendations received some attention worth mentioning, e.g.,~\citep{Angara2017,zeng2018eliciting,Tong:MM2019:VisionLanguage,Shimazu2002}. Research on other domains is, however, very sparse, including a number of application cases which were only addressed in one single paper, e.g., the recommendation of podcasts, scientific papers or car tires~\citep{Yang:2018:UUI:3240323.3240389,BotWheels2017,Loh2010}.

\subsection{Evaluation: Paradigms and Experimental Setups}
\begin{figure}[h!t]
  \centering
  \includegraphics[width=0.5\linewidth]{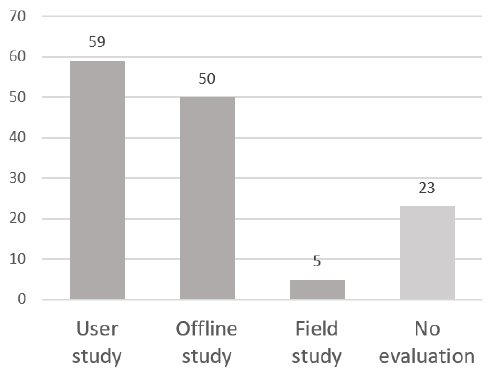}
  \caption{Evaluation paradigms.}
  \label{fig:landscape-evaluation-paradigms}
\end{figure}

Next, we analyze how researchers approach the problem of evaluating their systems. Figure~\ref{fig:landscape-evaluation-paradigms} shows how often the different evaluation paradigms from Section~\ref{sec:evaluation-methods} were applied. Like in Figure~\ref{fig:landscape-domains}, we count papers more than once in case they used a mixed evaluation methodology.

Figure~\ref{fig:landscape-evaluation-paradigms} shows that in about 45\% of the papers humans were involved in the evaluation process in some ways. A large fraction of the papers, about 40\%, included a computational analysis. In most of these cases, authors however relied on only one of these paradigms, i.e., in the large majority of cases, they either only used a user-centric evaluation or a computational analysis. Only 10\% of the papers combined two or more evaluation approaches, e.g.,~\citep{qiu2017alime,liao2019deep,zhao2019personalized,Christakopoulou:2018}. As a result, in about 30\% of all papers the proposed system or technical approach was evaluated entirely through computational experiments. Reports on field studies are generally rare, which is also the case in general recommender system research literature.
Finally, we observe that in about 18\% of the papers, no experimental evaluation was provided. This was for example the case for papers where mainly ideas or prototypes were presented.\footnote{Remember that in our analysis we did not include papers without a technical contribution in the first place, such as survey papers or papers that propose or analyze datasets for conversational recommendation.}

The number of people who are involved in user-centric evaluations typically depends on the type of method.
\begin{itemize}
\item In cases where human judges are recruited to assess or compare individual responses by different CRS, usually only a handful of human experts are involved, see e.g.,~\citep{li2018towards,chen-etal-2019-towards}. In some cases, little is said about the background or expertise of the human judges or about the specific instructions they received for the judgment task.
\item The number of participants in user studies (e.g., lab experiments) varies strongly, ranging from about 10-20 participants in early-stage research projects up to several hundred participants, e.g., in the case of the chatbot system proposed by~\cite{Narducci2018aiai}. Some of these larger studies, e.g.,~\citep{ashktorab2019resilient}, are conducted with the help of crowdworkers using platforms such as Amazon Mechanical Turk. Most commonly, user studies are however conducted involving a few dozen participants, depending on the research question and study design.
\item In the few field experiments reported in the analyzed papers, in many cases no exact numbers are provided regarding the number of users and papers merely mention that a fraction of the live traffic was diverted to the new system, e.g.,~\citep{Christakopoulou:2018}. Participant numbers are therefore only available in a few cases, e.g., when an academic partner was involved in the development of a CRS that was field tested, see, e.g., the study by~\cite{Mahmood:2009:IRS:1557914.1557930}, which involved several hundred online users in the evaluation of a travel-planning system.
\end{itemize}

Regarding the study designs used in user-centric research on CRS, all sorts of configurations can be found in the literature, including between-subjects and within-subjects experimental designs, as well as a number of studies where participants only interacted with one (the proposed) system and there was no control group, e.g.,~\citep{Hong:2010:IAS:2108616.2108681,warnestaal2005user}. In this latter case, the evaluation of the system is frequently based on standardized instruments such as the System Usability Scale, which is interpreted on an absolute scale. Other forms of non-experimental research (e.g., exploratory studies) are very rare in the literature.

Finally, looking at the \emph{targets} of the evaluation, we find that laboratory studies usually aim to assess the quality perception of a CRS \emph{as a whole}, including, e.g., dialogue quality, recommendation accuracy, UI aspects, or overall utility. Computational studies, in contrast, often focus on specific algorithmic components of the CRS, e.g., the method to determine suitable recommendations, the approach to decide on the next question to ask, or other components in the processing pipeline such as modules for intent classification, entity recognition, or sentiment analysis. % TODO: could add some references here and there.

\subsection{Evaluation Measures}
\label{subsec:evaluation-measures}
CRS can be evaluated along a variety of quality dimensions, as discussed in Section~\ref{subsec:quality-dimensions}. Here, we provide an overview of the predominant approaches in the research literature. As our main categorization scheme, we differentiate between subjective and objective measures.

\subsubsection{Subjective Measures}
\label{susbec:subjective-measures}
Subjective quality perceptions regarding one or more systems are in most cases assessed
with the help of questionnaires that participants fill in the context of a study, e.g., after performing a specific task with a CRS. Alternative forms of subjective evaluations exist as well, for example, where participants express their \emph{relative} preferences for one or the other system. Less structured approaches include methods for exploratory research, e.g., interviews or focus groups. In the case of questionnaires, these are either based on established research instruments---such SUS or the ResQue framework for recommender systems~\citep{Pu:2011:UEF:2043932.2043962}---or specifically designed for the purpose of the study. In most cases, such questionnaires cover more than one quality dimension.

\paragraph{A Catalog of Subjective Measurement Dimensions for CRS}
The analysis of the papers considered in this study revealed that a rich variety of subjective quality dimensions are used in the literature.
We organize the identified quality dimensions based on the ResQue framework, which itself is based on the Technology Acceptance Model (TAM)~\citep{TAM1989} and the Software Usability Measurement Inventory (SUMI)~\citep{SUMI93}. Furthermore, since a number of more recent CRS implemented as chatbots, we incorporate interaction-related structures of the evaluation framework for chatbots proposed by~\cite{radziwill2017evaluating} in our catalog.

\begin{figure}[h!t]
  \centering
  \includegraphics[width=0.99\linewidth]{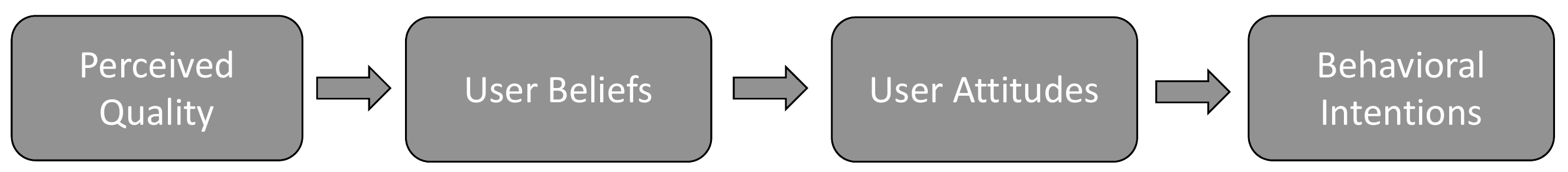}
  \caption{Overview of Main Concepts of the ResQue Model.}
  \label{fig:landscape-subjective-measures}
\end{figure}

Figure~\ref{fig:landscape-subjective-measures} provides an overview of the main elements of the ResQue model. The model generally assumes that various user-perceived \emph{quality factors}, e.g., recommendation accuracy,  may have an impact on thee \emph{users's beliefs} about the system, e.g., regarding its usefulness or transparency.  These believes may then influence user \emph{attitudes}, e.g., trust and satisfaction, ultimately potentially influencing their \emph{behavioral intentions}, e.g., to use the system again in the future. The ResQue model is widely applied in the recommender systems literature, and, due to its generality, it was also used in the context of the evaluation of conversational recommender systems~\citep{Jin:2019:MEC:3357384.3357923,Dietz2019,Pecune2019AmodelSocial,Alvarez2016}.

The model concepts on the right-hand side, user attitudes and behavioral intentions, are very general ones, and apply for any type of recommender system. For the elements that are more on the left-hand side, we however found a number of aspects regarding, e.g., different types of quality and usability, which are often specific to CRS.
In our categorization, we have therefore separated specific usability aspects from user beliefs. Furthermore, we have introduced sub-categories for different quality dimensions. Specifically, we differentiate between recommendation quality, conversation (dialogue) quality, and information quality factors. The resulting table of subjective quality measures is shown in Table~\ref{tab:subjective-measures}.

\small
\begin{longtable}{@{}p{-1cm}p{4cm}p{6.8cm}@{}}
\caption{Categorization of Subjective Quality Measures} \\
\multicolumn{2}{@{}l}{\textbf{Recommendation Quality}}                         & \textbf{Remarks}                                                                                                                                                                                                                       \\
                   & Accuracy                                               & \multirow{6}{6.8cm}{\justifying These quality factors are widely used and part of the ResQue model, which was used as one component for CRS evaluation in~\citep{Jin:2019:MEC:3357384.3357923,Dietz2019,Pecune2019AmodelSocial,Alvarez2016}. Various terms for accuracy are used in the literature,  e.g., \emph{match with user preferences}~\citep{DBLP:journals/expert/RicciN07}, \emph{perceived accuracy}~\citep{Contreras2018}, or \emph{recommendation quality}~\citep{Grasch:2013:RTC:2507157.2507161}. The \emph{attractiveness} of entire item sets was, e.g., considered in a non-CRS work in \citep{Willemsen2016Understanding}. \emph{Context compatibility} refers to the question of the recommended items match a user's current contextual situation.}   \\
                   & Attractiveness                                         &                                                                                                                                                                                                                                        \\
                   & Context compatibility                                  &                                                                                                                                                                                                                                        \\
                   & Diversity                                              &                                                                                                                                                                                                                                        \\
                   & Familiarity                                            &                                                                                                                                                                                                                                        \\
                   & Novelty                                                &                                                                                                                                                                                                                                        \\
                   &                                                        & \\
                   &                                                        & \\
                   &                                                        & \\
                   &                                                        & \\[+4em]
\multicolumn{2}{@{}l}{\textbf{Interaction   Quality}}                          &                                                                                                                                                                                                                                        \\
&  Affect & Affect is a main dimension for chatbot evaluation in~\citep{radziwill2017evaluating}. ~\cite{SinghcinICD102019} assess if their CRS agent ``provide a greeting'' or has a ``pleasant personality''. Various works also investigate if the interaction with a given CRS is \emph{engaging}, \emph{entertaining}, \emph{pleasant}, \emph{rewarding} or \emph{fun}, e.g.,~\citep{warnestaal2005user,LEE201795,DECAROLIS201787}\footnote{Sometimes, such qualities are considered part of the general usability of a system.}~\cite{Pecune2019AmodelSocial} furthermore consider categories like \emph{positivity}, \emph{rapport}, \emph{mutual attentiveness}, \emph{reciprocity}, and \emph{coordination}. Affect in Interaction Quality may impact user beliefs about \emph{Enjoyment}, see below.\\ [0.7em]

& Flow / Humanity                  & Another main dimension for chatbot evaluation from~\citep{radziwill2017evaluating}.~\cite{SinghcinICD102019}, e.g., investigate if their CRS is ``able to maintain a themed discussion'' and ``able to respond to specific questions''.~\cite{moon2019opendialkg} assess the relative \emph{naturalness}, see also~\citep{crsal2020} and \emph{relevance} of generated dialogue transitions.~\cite{chen-etal-2019-towards} let human judges score the \emph{consistency} of given utterances with the dialogue history.~\cite{zhangUserSimulation2020} ask judges which of two dialogues was performed by a real user. \\ [0.7em]
& Input Processing Performance                             & Related measures proposed in the literature are \emph{intent detection accuracy}~\citep{Cultural2019}, \emph{interpretation performance}~\citep{warnestaal2005user}, \emph{comprehension}~\citep{crsal2020} or \emph{robustness (to unexpected input)}~\citep{SinghcinICD102019}. Intent detection accuracy was measured in~\citep{Cultural2019} by directly asking the users of a chatbot solution about this aspect. In~\citep{warnestaal2005user}, interpretation performance is defined as: ``The user's experience of how well the system understands her input.'' \\

& Interaction adequacy  &  Relates to the availability of suitable interaction mechanisms for, e.g., preference expression, preference revision and also explanation~\citep{Pu:2011:UEF:2043932.2043962}.  \\
                   \\[-0.7em]
\multicolumn{2}{@{}l}{\textbf{Information   Quality}}                          &                                                                                                                                                                                                                                        \\
& Adaptation                                             & Used in~\citep{warnestaal2005user,LoeppChoiceBased2014} to assess if the system adapts to user preferences; overlaps also with recommendation quality.~\cite{FuzzyBased2018} use the term ``Appropriateness'', expressing if the content is varied according to user proficiency.~\cite{WALKER2004811} investigated affects of the adaptation of various factors such as \emph{conciseness} (related to \emph{comprehensibility}) and the \emph{mode} in which information is presented. Generally, adaptation is related to personalization as a usability dimension, e.g., in~\citep{zhao2019personalized}. \\
& Coherence \& correctness &  The general coherence and correctness of a chatbot's responses was used as criterion in~\citep{Cultural2019}.
\\
& Fluency                                                & Human judges were asked to assess the fluency of the generated responses by the system in~\citep{liao2019deep} \and{\citep{zhao2019personalized}}. A precise definition of fluency is not always provided. \\
& Information sufficiency                                & Captures if the information provided by the CRS, e.g., about the items, is sufficient to ``\emph{facilitate users' decision making processes}''~\citep{PuCH12Evaluating}. This measure is part of the \emph{interface adequacy} concept in the ResQue framework.~\cite{warnestaal2005user} uses the related term \emph{domain coverage} to measure to what extent users feel that there are enough items to chose from and if there is enough information about the items provided by the system.
 \\
& Informativeness                                        & Human judges were asked to assess the informativeness of the generated responses by the system in~\citep{liao2019deep} and~\cite{kgsf2020}; informativeness for example covers if user queries were properly answered. %A more detailed definition of the concept is not provided in~\citep{liao2019deep}.
This measure is therefore related but different from \emph{information sufficiency}, which is focused about information regarding the items. \\
& Quality of returned responses                          & Domain experts were asked to assess the general quality of the auto-generated responses in~\citep{FuzzyBased2018}.  \emph{Generation performance} was used as a quality dimension in~\citep{warnestaal2005user}, covering phrase choice, clarity, and verbosity. \\% \emph{Text quality}~\citep{zhao2019personalized} \\
& \\[-0.7em]
\textbf{Usability} &                                                        &                                                                                                                                                                                                                                        \\
                   & Aggregated usability                              & A number of research works rely on general usability assessment instruments for software such as the System Usability Scale (SUS) and pre-defined questionnaire items, e.g.,~\citep{MapBasedRicci2010,Grasch:2013:RTC:2507157.2507161,Alvarez2016}. Note that the SUS instrument, which results in a number that can be interpreted on an absolute scale, covers several aspects that are analyzed individually in other papers, e.g., ease-of-use or intention-to-use in the future.\\
                   & Ease-of-use                                       &  Ease-of-use was assessed as an usability dimension in different works, e.g.,~\citep{Contreras2014}, usually in combination with other factors and when the Technology Acceptance Model (TAM)~\citep{TAM1989} is used~\citep{DBLP:conf/recsys/PuZC09}. Sometimes, ease-of-use is seen related or equated to \emph{efficiency} and \emph{effort}. \\
                   & Interface adequacy &  This quality dimension was proposed for the ResQue model and is, for example, ``\emph{concerned with how to optimize the recommender page
layout to achieve the maximum visibility of the recommendation}'' \citep{Pu2010UcerstiUser}. More general guidelines regarding the user interface design of recommenders are provided, e.g., in~\citep{OzokDesign2010}.\\
                   & Learnability                             & Part of the SUS questionnaire\footnote{E.g.:``\emph{I would imagine that most people would learn to use this system very quickly.}''}, but sometimes assessed individually, e.g., in~\citep{Contreras2018}; sometimes referred to as \emph{easy-to-learn}. \\
                   & Responsiveness                                         & An assessment regarding how fast the system responds; a measure of interaction pace~\citep{warnestaal2005user}. Also considered in~\citep{Clarizia2018}.\\
                   &                                                        & \\[-0.7em]
                   \multicolumn{2}{@{}l}{\textbf{User   Beliefs}}                                 &                                                                                                                                                                                                                                        \\
\textbf{}          & Aggregated System Quality (general)                               & In some works like~\citep{Pecune2019AModel,ResearchNoteContingencyApproach2013} overall system quality is assessed through aggregating user perceptions in different dimensions based, e.g., on the ResQue model or on the trust model from~\citep{DBLP:journals/isr/McKnightCK02}.\\
                   & Collaboration                                          & The extent to which a CRS in a virtual environment supported collaboration between users was assessed in~\citep{Contreras2014,Contreras2018}. \\
                   & Control                                                & To assess if users feel in control of the selection item process, used in~\citep{LoeppChoiceBased2014}. \\
                   & Expected behavior                                      & Considered as one of several aspects leading to satisfaction in~\citep{warnestaal2005user} and measures ``\emph{how intuitive and natural the dialogue interaction is.}'' \\
                   & Enjoyment                                              & Used in a variety of works, e.g.,~\citep{FactorsInfluencing2021,Contreras2014,warnestaal2005user}. Related factors are entertainment, pleasantness, rewardingness, ``liking the interaction'', fun, or engagingness. \\
                   & Transparency                                           & A main construct in the ResQue framework; mostly refers to \emph{perceived transparency} and how users think about the inner logic of a system. Related to \emph{control} and \emph{trust}~\citep{LoeppChoiceBased2014}.\\
                   & Usefulness                                             & Beliefs about the extent a system helps users accomplish their (selection) task, e.g., in~\citep{MapBasedRicci2010,FactorsInfluencing2021,crsal2020} and works based on the TAM model. Related also to \emph{task ease}~\citep{warnestaal2005user}. \\
                   &                                                        &                                                                                                                                                                                                                                        \\ [-0.7em]
                   \multicolumn{2}{@{}l}{\textbf{User   Attitudes}}                               &                                                                                                                                                                                                                                        \\
\textbf{}          & Confidence                                             & \multirow{4}{6.8cm}{Confidence \& Trust  and Satisfaction are proposed in the ResQue model, based on the TAM model.~\cite{DBLP:journals/jmis/XuBC17} used the term ``decision quality'' for consumers' decision confidence.} \\
                   & Satisfaction                                           &                                                                                                                                                                                                                                        \\
\textbf{}          & Trust                                                  &                                                                                                                                                                                                                                        \\
                   &                                                        &                                                                                                                                                                                                                                        \\[1em]
\multicolumn{2}{@{}l}{\textbf{Behavioral   Intentions}}                        &                                                                                                                                                                                                                                        \\
                   & Intention to reuse, future use                        & \multirow{3}{6.8cm}{Part of the ResQue model; used also in works that adopt the TAM model, e.g.,~\citep{Contreras2014} or~\cite{fadhil2019assistive}. Social intentions may for example be to share or give feedback on social media.}  \\
                   & Recommend to friend, social intentions                 & \\
                   & Intention to purchase                                               &
\label{tab:subjective-measures}
\end{longtable}
\normalsize

Generally, with the ResQue model, we rely on a conceptual framework in this section that is widely adopted in the relevant literature. Moreover, the model is specifically tailored to user-centric research approaches and allows us to organize existing approaches in a more fine-grained way than with four high-level quality dimensions introduced in Section~\ref{sec:evaluation-methods}. %from~\citep{jannach2021crscsur}.

The subjective measures listed in Table~\ref{tab:subjective-measures} can however be mapped to the four dimensions discussed in Section~\ref{sec:evaluation-methods}. The measures listed under \emph{Recommendation Quality} in Table~\ref{tab:subjective-measures}, for example, mainly correspond to the dimension \emph{Effectiveness of Task Support} introduced in~\citep{jannach2021crscsur}. Measures in the dimensions \emph{Interaction Quality}, \emph{Information Quality}, and \emph{Usability} can be largely mapped to \emph{Quality of the Conversation and Usability}. The mapping between the categorization schemes is however not always one-to-one. In some usability measurements, for example, questions of efficiency, ease-of-use and responsiveness are included, which would be mapped to \emph{Efficiency of Task Support} in the alternative categorization scheme. Likewise, some measures listed under \emph{User Beliefs} in the ResQue framework, e.g., Control, would probably fall under \emph{Quality of the Conversation and Usability}, whereas the Usefulness measure rather relates to \emph{Effectiveness of Task Support}. The dimensions \emph{User Attitudes} and \emph{Behavioral Intentions}, finally, may be seen as consequences of \emph{Effectiveness of Task Support}. We observe that only few subjective measures relate to the \emph{Effectiveness of Subtask} dimension. This is expected as user studies often focus on the overall perception of the system, e.g., in terms of usability and recommendation quality, but not on internal specifics such as Named Entity Recognition. One exception is the Input Processing Performance measure listed under \emph{Interaction Quality}, which explicitly considers the aspect of intent-recognition accuracy, which is a common subtask in many CRS.

\paragraph{Subjective System Comparisons}
In most studies with users that use subjective measures, the study participants interact with one (within-subjects) or more (between-subjects) variants of a CRS and then report their perception of the system in one or more of the dimensions shown in Table~\ref{tab:subjective-measures}. Independent of the particular design, the responses are typically reported on an absolute scale, e.g., on a range from 1 to 5.

However, in a few studies, also \emph{relative} judgments are collected, where human evaluators express a preference for one of several systems or provide a ranking among the available systems, e.g.,~\citep{li2018towards,chen-etal-2019-towards,crsal2020,ashktorab2019resilient,zhangUserSimulation2020,hayati-etal-2020-inspired,zhang2021kecrs,manzoorrs2021}. In such studies, often a small set of human evaluators are involved, and sometimes they have certain competencies required for the task, e.g., linguistic knowledge~\citep{chen-etal-2019-towards}.

In~\citep{li2018towards} for example, ten human evaluators were tasked to rank the responses of three systems in several dialogue situations according to their \emph{overall quality}. In~\citep{chen-etal-2019-towards}, human judges had to assess the responses of three systems according to their \emph{consistency with the dialogue}, using an absolute scale from 1-3. \cite{manzoorrs2021} later on conducted a reproducibility study where three systems were evaluated with respect to the ``meaningfulness'' of their responses. The judges in the work by~\cite{zhangUserSimulation2020} were asked to tell which of two presented dialogues were performed by humans.
For the commercial AliMe Chat system~\citep{qiu2017alime}, business analysts were involved who graded the responses of the proposed system and another existing system using a three-point scale. Five criteria were provided for evaluating the responses by the query-answering system: were the responses ``right in grammar'', ``semantically related'', in ``well-spoken language'', ``context independent'', and ``not overly generalized''.

A pairwise comparison was also done in~\citep{ashktorab2019resilient}, where study participants were shown pairs of possible conversation \emph{repair} scenarios, and they were asked which \emph{``which scenario appealed to
them more and describe why they had made their selection.''} Differently from previous works, the evaluation was not focused on general system responses, but on the specific ways a system is dealing with conversational breakdowns.

\subsubsection{Objective Measures}
\label{subsec:objective-measures}

Objective measures are those that are collected or computed automatically. Such measures can be used both for studies involving users, e.g., to measure the time study participants need, and for simulation/offline studies, e.g., to assess the accuracy of the recommendations made by the system. Typically, objective measures are used to assess a particular aspect of a given CRS, e.g., the efficiency of the conversation in terms of observed interaction cycles or if the user clicked on one of the recommendations.

\paragraph{A Catalog of Objective Measures for CRS} A variety of objective measures is used in the surveyed literature. In
Table~\ref{tab:objective-measures}, we provide an overview of these measures. We organize the measures according to the four dimensions proposed in~\cite{jannach2021crscsur} and summarized in  Section~\ref{subsec:quality-dimensions}:
\begin{itemize}
  \item \emph{Effectiveness of Task Support}, i.e., the capability of a system to achieve its main task, e.g., helping users making a decision or finding something relevant;
  \item  \emph{Efficiency of Task Support}, i.e., helping users to make a decision or find something relevant with limited effort or in short time;
  \item \emph{Conversation Quality and Usability}, including, e.g., aspects relating to linguistic properties of the dialogue,  and
  \item \emph{Effectiveness of Subtask}, e.g., in terms of Named Entity Recognition accuracy.
\end{itemize}

\small
\begin{longtable}{@{}p{-1cm}p{4cm}p{6.8cm}@{}}
\caption{Categorization of Objective Quality Measures} \\

\multicolumn{2}{@{}l}{\textbf{Effectiveness of Task Support}}                          & \textbf{Remarks} \\

& Accuracy                          & A variety of accuracy measures from Machine Learning and Information Retrieval are used for offline experiments including, Precision, Recall, F1, AUC, Hit Rate, RMSE, MAE, NDCG, Reward/Regret.
In user studies, alternative quantitative measures are used, e.g., the ``position of the selected item'' or the ``fraction of users switching their decision later on''~\citep{4641338,Contreras2018}. \\
& Adoption & In user studies, \emph{task completion rates}, \emph{success rates} or \emph{(virtual) purchases} can inform about the adoption of the recommendations or advice by users, e.g.,~\citep{ResearchNoteContingencyApproach2013,Mahmood:2009:IRS:1557914.1557930,Sun:2013:LMD:2433396.2433451,vision2019,tsumita2019dialogue}. In live studies, for example click-through-rates or the number of users opening notifications can be measured~\citep{zhao2019personalized,Christakopoulou:2018} \\
& Engagement & In contrast to the efficiency perspective, more interactions are sometimes considered a signal of engagement and success of a recommender. Engagement can, e.g., be measured in the number of listened songs, video watch time, or stay time~\citep{Christakopoulou:2018,Jin:2019:MEC:3357384.3357923,Kamej2010} \\
& \\[-0.7em]
                   &                                                        & \\[-0.7em]
\multicolumn{2}{@{}p{4cm}}{\textbf{Efficiency of Task Support}}                         &  \\
                   & Choice set reduction & Some systems aim to reduce the number of remaining options during the conversation to increase efficiency. Related terms are the \emph{number of unique cases presented}, \emph{result set size}, \emph{remaining items}, \emph{number of cases to inspect} or \emph{pruning rate}, e.g.,~\citep{Shimazu2002,Rafter2005,Trabelsi2013,AICS:03:Smyth:FeedBStratsConvRecSys} \\
                   & Interaction counts & The number of needed \emph{interaction cycles} is widely used in the literature, both for simulation and user studies. Other, less frequently used measures include the number of \emph{clicks}, or \emph{inspected items} (viewed/listened/watched), e.g.,~\citep{DynamicPers2014,Jin:2019:MEC:3357384.3357923,Dietz2019}. \\
                   & Time Measurements & Task completion times (or: session times) are commonly used in user studies to inform about the time needed until a recommendation is found. In some cases, computation and running times are reported for simulation experiments~\citep{Llorente2012,Trabelsi2013}.\\
                   &                                                        & \\[-0.7em]
\multicolumn{2}{@{}l}{\textbf{Conversation Quality }}                                 & \\

& Linguistic Properties & Various measures are applied to gauge the quality of the system-generated utterances, e.g., sentence-level and corpus-level BLEU, perplexity, lexical diversity, distinct n-gram, NIST, or ROUGE~\citep{nie2019multimodal,Greco2017,DBLP:conf/aaai/GhazvininejadBC18,chen-etal-2019-towards,crsal2020} \\

& Usefulness of Interaction & Different assessments are made in~\citep{Cerezo2019} to gauge if users were generally \emph{able to interact} with a chatbot. Other works examine more specific aspects like the frequency of users performing certain actions (e.g., compound critiques) or the fraction of preferences that could be discovered in the dialogue~\citep{tsumita2019dialogue,McCarthy2004Onthe,baizal2017,DBLP:conf/recsys/ViappianiPF07}. \\[1em]

\multicolumn{2}{@{}l}{\textbf{Effectiveness of Subtask}}                        & \\
                   & Analysis of Dialogue Situation & Typical tasks include intent recognition/classification (and utterance selection) or the detection of chit-chat situations, which can be assessed with classification accuracy measures, e.g.,~\citep{nie2019multimodal,Thompson:2004:PSC:1622467.1622479,tsumita2019dialogue,IOVINE2020113250,DBLP:conf/aaai/YanDCZZL17}. \\
                   & Input Processing & Entity recognition is a common subtask in CRS; more specific problems are category detection or keyphrase detection. Accuracy measures, success rates, or task-specific measures can be applied~\citep{Wu2019,liao2019deep,DBLP:conf/aaai/YanDCZZL17}\\\
                   & Retrieval and Analysis of Content & \cite{nie2019multimodal} for example proposed a multimodal system and measured Recall for the task of selecting  images to show to users. \cite{li2018towards} evaluated the performance of the sentiment analysis component of their system. \\
\label{tab:objective-measures}
\end{longtable}
\normalsize
\section{Discussion}
\label{sec:discussion}
Our survey reveals a major trend from more traditional form-based systems towards (end-to-end) learning approaches to build CRS that support natural language interaction. These developments however also render the evaluation of modern CRS more challenging. In particular, these novel techniques require appropriate mechanisms to assess the quality both of individual system utterances and the resulting recommendation dialogue as a whole. As our survey shows, a number of alternative ways of evaluating such aspects were explored in recent years, using a multitude of objective and subjective measures. Still, various challenges regarding the evaluation of CRS remain.

\paragraph{Understanding User Expectations and Realistic Datasets.} First, it seems that more foundational research is needed to understand what would be considered desirable properties of a conversational recommendation agent. What are the expectations by users? Which intents should be supported? For example, should the system be able to explain its recommendations? How important is chit-chat? Do users really expect human-like performance or would they be satisfied with a system that fails from time to time to understand user utterances?

Only very few works in the surveyed literature approach the problem of building a CRS by first analyzing certain fundamentals, for example, how humans interact in a recommendation dialogue. Understanding such ``natural'' behavior could however be one main source of knowledge that informs the design of a computerized recommender. While there are a few works that examine existing datasets in terms of, e.g., the role of ``sociable'' recommendation strategies~\citep{hayati-etal-2020-inspired} or  chit-chat, we found no work that actually examined the expectations by humans regarding the capabilities of a CRS.

One hypothesis in the context of modern learning-based approaches could be that such knowledge is not needed, because the system will be able to learn at some stage how to respond appropriately when only given a sufficiently large amount of training data. Today's datasets for learning, e.g., the ReDial dataset, however seem to be too limited to allow for such a learning approach. While the number of recorded dialogues is not small, a closer look at the corpus reveals that many of these dialogues only use a narrow set of dialogue patterns. Questions about \emph{why} a certain movie was recommended are for example rarely asked in this dataset. This might be caused by the specific instructions given to the crowdworkers that were used to develop the corpus. As a result, even the best learning technique will face limitations when it is trained one such somewhat artificial dialogues. Ultimately, this also calls for the creation of new datasets that better reflect the richness of human-to-human conversations. The INSPIRED dataset proposed in~\citep{hayati-etal-2020-inspired} represents an important step in that direction, as it shows that more sociable behavior in reality more often leads to recommendation success. Yet another dataset, named \emph{DuRecDial}, for more natural conversations was proposed by~\citep{liu-etal-2020-towards}. Here, a dialogue can both have non-recommendation parts and dialogue turns about recommendation, and the goal of a chatbot can be to lead users to a recommendation dialogue. As an objective measure, the authors also propose to use the level of \emph{proactivity} of the chatbot. Related considerations also inspired the development of a dataset for ``topic-guided'' dialogues by~\cite{zhou-etal-2020-towards}.

Another line of future research may also aim towards alternative ways of acquiring user preferences in a more natural way. Today, ``slot-filling'' is still a predominant approach, where the main assumption is that users will be able to specify their needs in terms of particular attributes of an item, e.g., its prize. In real-world conversations, and in particular in sales situations, a human recommendation agent would probably often rather ask about the \emph{desired functionality} or the \emph{intended use} of a specific item. Such considerations can be implemented both in traditional interactive sales advisory solutions~\citep{widyantoro2014framework,Jannach:2004:ASK:3000001.3000153}, but also in modern approaches based on machine learning. In a recent work \cite{Kostric2021Soliciting} present a promising approach for generating what they call \emph{implicit questions}, which may help advance natural-language based systems beyond today's predominant slot-filling approaches. Furthermore, \cite{Radlinski2022SubjectiveAttributes} investigate challenges of understanding and modeling \emph{subjective attributes} to overcome limitations of traditional slot-filling techniques.

\paragraph{The Need for Mixed-Method Approaches.}
Our survey shows that a considerable number of papers, about 30\%, are exclusively based on offline experimentation. In some other papers, the main results are also based on computational analyses, but complemented with a study involving humans. Unfortunately, the user-centric part is sometimes very brief, with little information provided about, e.g., how the human judges were selected or what the specific instructions for the judges were.

While offline experimentation can be helpful to assess certain aspects of a CRS like speech recognition accuracy, offline evaluations often come with numerous limitations. Even for the commonly-studied problem of predicting the relevance of items for users some questions remain. In the end, it is not clear if better results in offline experimentation actually leads to systems that are perceived as more helpful by users or systems that are better in terms of the provider's goal~\citep{jannach2021mcnamara}. Objective evaluations of the linguistic quality of generated system responses are challenging as well. Metrics like the BLEU score are, for example, not undisputed as a means for evaluation~\citep{liu-etal-2016-evaluate} and it is unclear if such measures actually reflect quality perceptions of users.

As a result, we argue that more often a mixed-methods approach should be followed, with a strong focus on user-centric research. In that context, we also believe that more \emph{exploratory} research is needed, in particular to understand the needs and expectations of CRS users, as discussed above. Various exploratory research methods actually exist, but these are only used infrequently. As a consequence, research on CRS more often requires teams that have expertise in different subareas of computer science. For example, building and evaluating a CRS can barely be done based on machine learning expertise alone, but may in many cases require the involvement of experts from fields like human-computer interaction.

A positive observation in that context is that many recent works on end-to-end learning have a human evaluation component besides the offline evaluation. Often, this is done by asking a set of human judges to assess individual responses of different systems, e.g., with respect to fluency and informativeness. While such studies are useful to compare systems, it does not tell us if the individual systems are close to being useful in practice. An analysis of two recent end-to-end learning systems in~\cite{JannachManzoor2020} revealed that about one third of the responses returned by the systems were not considered meaningful by independent evaluators. With such a high error rate, these systems will inevitably lead to broken conversations at some stage. More research is therefore also required to understand when such systems fail.

\paragraph{Towards a More Standardized Evaluation Approach}
In the area of recommendation \emph{algorithms}, researchers have developed a quite standardized research approach over the years, which adopts principles from the fields of information retrieval and machine learning. For the area of user-centric \emph{systems} evaluation, also proposals for general evaluation frameworks were made, in particular by~\cite{Pu:2011:UEF:2043932.2043962} and~\cite{umuai2012knijnenburg}. These general frameworks for user-centric recommender systems research are only used to a certain extent in today's CRS research. In a few works, researchers furthermore rely on the very general System Usability Scale for their evaluation. More frequently, however, researchers design their own measurement methods and instruments (questionnaires), often based on parts of existing frameworks or the specific needs of their research question. This makes the comparison of works by different researchers and thus, the assessment of progress, difficult.

The development of a more standardized research and evaluation approach is therefore desirable, see also the efforts for the more general problem class of task-oriented dialogue systems~\citep{finch-choi-2020-towards}.
While existing frameworks like ResQue can in principle be applied, these frameworks do not account for the specifics of CRS yet. Future work might therefore aim at the extension of these more general frameworks, e.g., with instruments that were developed for the evaluation of chatbots. As many modern CRS are implemented in the form of a chatbot, one may apply the corresponding quality criteria, e.g., if the system is engaging, error tolerant, able to deal with conversation breakdowns, or avoids inappropriate language.

Looking at objective measures, the measures for recommendation accuracy are widely standardized (e.g., RMSE, precision, recall etc.). For other aspects, in particular for linguistic qualities, no clear standard exists yet. However, some of the measures seem to be more widely used than others. In an effort to harmonize offline evaluation, researchers recently proposed a common evaluation framework CRSLab~\citep{zhou2021crslab}, which features a number of baseline algorithms and evaluation measures such as BLEU or distinct n-gram. Furthermore, the framework is designed to be able to process several of today's frequently used dialogue datasets.

Another CRS evaluation framework based on user simulation was recently proposed in~\citep{zhangUserSimulation2020}. One main goal of their work is to avoid the need for time-intensive and difficult human evaluation as far as possible. In their framework, users are simulated that generate responses like a human would probably do, and their experiments indicate that the approach is realistic and that there is a good correlation between human judgements and the implemented objective measures.

Despite these positive developments, we still observe a wide range of different evaluation approaches, both in the context of offline evaluations and evaluations with users. Based on a survey of the existing literature,~\cite{finch-choi-2020-towards} recently proposed a set of eight dimensions to evaluate the quality of dialogues in the context of task-oriented dialogue systems. Given that CRS represent an important subclass of such task-oriented systems, researchers should in the future more often consider these established quality dimensions as well in their evaluations.

In the broader context of standardized evaluations, related fields like Information Retrieval or Natural Language Processing have a longer tradition of relying on \emph{benchmarks} (test collections) consisting of ``shared tasks'' and datasets as a means to achieve and demonstrate progress. Today, no standardized or broadly used benchmarks exist for CRS. Experiences from related fields show that standardized benchmarks can be strong drivers of progress. However, benchmarks may also lead to a certain hyperfocus on a small number of tasks using a limited set of computational metrics, e.g., accuracy metrics for the recommendation task and linguistic metrics for dialogue quality. In the end, it may then remain unclear if small improvements on benchmarks would matter in practice. For CRS, this is a particularly pronounced problem, since conversational recommendation is a complex interactive process. Any holistic evaluation may at the end require a human in the loop to some extent in the evaluation process. For certain aspects, however, benchmarks may be helpful as they also foster the creation and sharing of new datasets, e.g., to develop novel machine learning approaches.

\paragraph{Provider Value and Multi-Stakeholder Evaluation}
Finally, little is reported in the literature about the value of CRS for providers. For more traditional, non-conversational recommender systems, various case studies exist in the literature~\citep{jannachjugovactmis2019}. In the area of CRS, however, only a few papers report of real-world deployments. Unfortunately, these descriptions are often very brief. It therefore remains largely unclear in which ways and to what extent CRS create utility both for users and providers in practice. Moreover, in traditional settings, the consideration of multiple stakeholders in the recommendation process has moved in the focus of research in recent years~\citep{abdollahpouri2020}. Similar efforts in the area of conversational approaches are unfortunately still missing.

\section{Conclusion}
Research in CRS has re-gained increased research interest in recent years due to various technological developments. Evaluating such complex, interactive systems, which commonly consist of a number of non-trivial components, can however be challenging. In this work we have reviewed the various quality dimensions of CRS and provided an overview on the various evaluation measures that are used in the literature. Our review in particular emphasizes the importance of human-in-the-loop evaluation approaches and certain limitations of today's research practices. Overall, we see our work as a step towards more holistic---and thus more realistic---evaluation practices in CRS research.

\section*{Acknowledgement}
I thank Ahtsham Manzoor for his valuable feedback during the creation of this manuscript.

\begin{comment}
\section*{Declarations}
\emph{Funding:} Not applicable. \\
\emph{Conflicts of Interest:}  Not conflicts of interest. \\
\emph{Availability of Data and Material:}Not applicable. \\
\emph{Code Availability: } Not applicable. \\
\emph{Author's Contribution:} Not applicable. \\
\end{comment}

\bibliographystyle{spbasic}      % basic style, author-year citations
%\bibliography{references}   % name your BibTeX data base

\end{document}